\documentclass{cup-hpl}

\usepackage{lipsum}
\usepackage[german, english]{babel}
\usepackage[utf8]{inputenc} 
\usepackage{amsmath}
\usepackage{amsfonts}
\usepackage{amssymb}
\usepackage{amsthm}
\usepackage{amsbsy}

\usepackage{graphicx}		
\usepackage{sidecap}
\usepackage{here}
\usepackage{chngcntr}
\counterwithin{figure}{section}
\usepackage[automark]{scrpage2}
\usepackage{geometry}
\usepackage{xcolor}

\usepackage{subcaption}
\usepackage{multirow} 
\usepackage{array} 
\usepackage{tabularx} 

\usepackage{verbatim} 

\usepackage{emptypage}

\begin{document}

\newtheorem{theorem}{Theorem}

\shorttitle{Polarized proton beams from laser-induced plasmas}                                   
\shortauthor{Anna Hützen \textit{et al.}}

\title{Polarized Proton Beams from Laser-induced Plasmas}

\author[1,2]{Anna Hützen \corresp{
		\email{a.huetzen@fz-juelich.de}}}
\author[3]{Johannes Thomas}
\author[4]{Jürgen Böker}
\author[5]{Ralf Engels}
\author[4]{Ralf Gebel}
\author[4,6]{Andreas Lehrach}
\author[3]{Alexander Pukhov}
\author[7,8]{T. Peter Rakitzis}
\author[7,8]{Dimitris Sofikitis}
\author[1,2]{Markus Büscher}

\address[1]{Peter Grünberg Institut (PGI-6), Forschungszentrum Jülich, Wilhelm-Johnen-Str.\,1, 52425 Jülich, Germany}
\address[2]{Institut für Laser- und Plasmaphysik, Heinrich-Heine-Universität Düsseldorf, Universitätsstr.\,1,
	40225 Düsseldorf, Germany}
\address[3]{Institut für Theoretische Physik I, Heinrich-Heine-Universität Düsseldorf, Universitätsstr.\,1,
	40225 Düsseldorf, Germany}
\address[4]{Institut f\"{u}r Kernphysik (IKP-4), Forschungszentrum J\"{u}lich, Wilhelm-Johnen-Str.\,1, 52425 J\"{u}lich, Germany}
\address[5]{Institut f\"{u}r Kernphysik (IKP-2), Forschungszentrum J\"{u}lich, Wilhelm-Johnen-Str.\,1, 52425 J\"{u}lich, Germany}
\address[6]{JARA-FAME und III. Physikalisches Institut B, RWTH Aachen, Otto-Blumenthal-Str., 52074 Aachen, Germany}
\address[7] {Department of Physics, University of Crete, 71003 Heraklion-Crete, Greece}
\address[8] {Institute of Electronic Structure and Laser, Foundation for Research and Technology-Hellas,
71110 Heraklion-Crete, Greece}

\begin{abstract}
We report on the concept of an innovative source to produce polarized proton/deuteron beams of a kinetic energy up to several GeV from a laser-driven plasma accelerator. Spin effects have been implemented into the PIC simulation code VLPL to make theoretical predictions about the behavior of proton spins in laser-induced plasmas. Simulations of spin-polarized targets show that the polarization is conserved during the acceleration process. For the experimental realization, a polarized HCl gas-jet target is under construction using the fundamental wavelength of a Nd:YAG laser system to align the HCl bonds and simultaneously circular polarized light of the fifth harmonic to photo-dissociate, yielding nuclear polarized H atoms. Subsequently, their degree of polarization is measured with a Lamb-shift polarimeter. The final experiments, aiming at the first observation of a polarized particle beam from laser-generated plasmas, will be carried out at the 10\,PW laser system SULF at SIOM/Shanghai.

\end{abstract}


\maketitle

\section{INTRODUCTION}
Ion acceleration driven by super-intense laser pulses has undergone impressive advances in recent years. Due to increased laser intensities, much progress in the understanding of fundamental physical phenomena has been achieved \cite{bib1,bib2,bib3,bib4}. Nevertheless, until today large-scale ion accelerators are used worldwide for producing energies up to 100\,MeV: from basic research, through semiconductor doping and isotope production, right up to medical applications, e.g., more efficient cancer treatment \cite{bib5}. However, appropriate accelerators such as cyclotrons, tandems, linear accelerators as well as storage rings are quite large, very energy-intensive and expensive in purchase and maintenance. \\ \hspace*{5 mm}
Laser-driven acceleration offers one highly promising alternative thanks to advances in laser technology. Increasing energies and repetition rates allow even higher ion energies and intensities, possibly even laser-induced nuclear fusion. In this context, one important feature of modern accelerators is still missing, namely the production of highly polarized particle beams. To achieve this, we are pursuing two approaches. First, polarization build-up by the laser itself and, second, polarization preservation of polarized targets during laser acceleration. Given that, one unsolved problem is the influence of the huge magnetic fields present in the plasmas acting on the ion spins. The present work aims at the first production of a polarized proton beam -- where the proton spins are aligned relative to an arbitrary quantization axis -- from laser-induced plasmas using polarized targets. \\ \hspace*{5 mm}
Two scenarios are discussed to build up a nuclear polarization in the plasma. Either polarization is generated due to a spin flip according to the Sokolov-Ternov effect by changing the spin direction of the accelerated particles, induced by the magnetic fields of the incoming laser pulse. Apart from that, the spatial separation of various spin states due to magnetic-field gradients (Stern-Gerlach effect) may result in the generation of polarization for different beam trajectories \cite{bib6}. \\ \hspace*{5 mm}
Besides these two mechanisms which may lead to a temporal or spatial polarization build-up, all particle spins precess around the laser or plasma magnetic fields as characterized by the Thomas-Bargmann-Michel-Telegdi (T-BMT) equation describing the spin motion in arbitrary electric and magnetic fields in the relativistic regime. \\ \hspace*{5 mm}
The first and only experiment measuring the polarization of laser-accelerated protons has been performed at the ARCturus laser facility at Heinrich-Heine University Düsseldorf \cite{bib2}. Figure\,1.1 schematically depicts the setup: for the measurements a 100\,TW Ti:Sa laser system with a typical pulse duration of 25\,fs and a repetition rate of 10\,Hz was used producing an intensity of several $10^{20}$\,Wcm$^{-2}$ when being focused on a target. Impinging the laser pulse in a 45$^{\circ}$ angle on an unpolarized gold foil of 3\,µm thickness, protons with an energy of typically a few MeV are produced. They are accelerated according to the Target Normal Sheath Acceleration (TNSA) mechanism \cite{bib1} towards a stack of three Radio-Chromic-Film (RCF) detectors where the number of protons is measured. In a Silicon target with a thickness of 24\,µm elastic scattering takes place. Thus, the spin-dependent asymmetries of the differential cross-section for the different azimuthal angles can be measured by counting the number of colliding particles per detector area with the help of CR-39 detectors that are placed a few millimeters downstream. The result was that no polarization was built up in the laser-accelerated proton beam.
 
\begin{figure}[H]
\centering
\includegraphics[width=80mm]{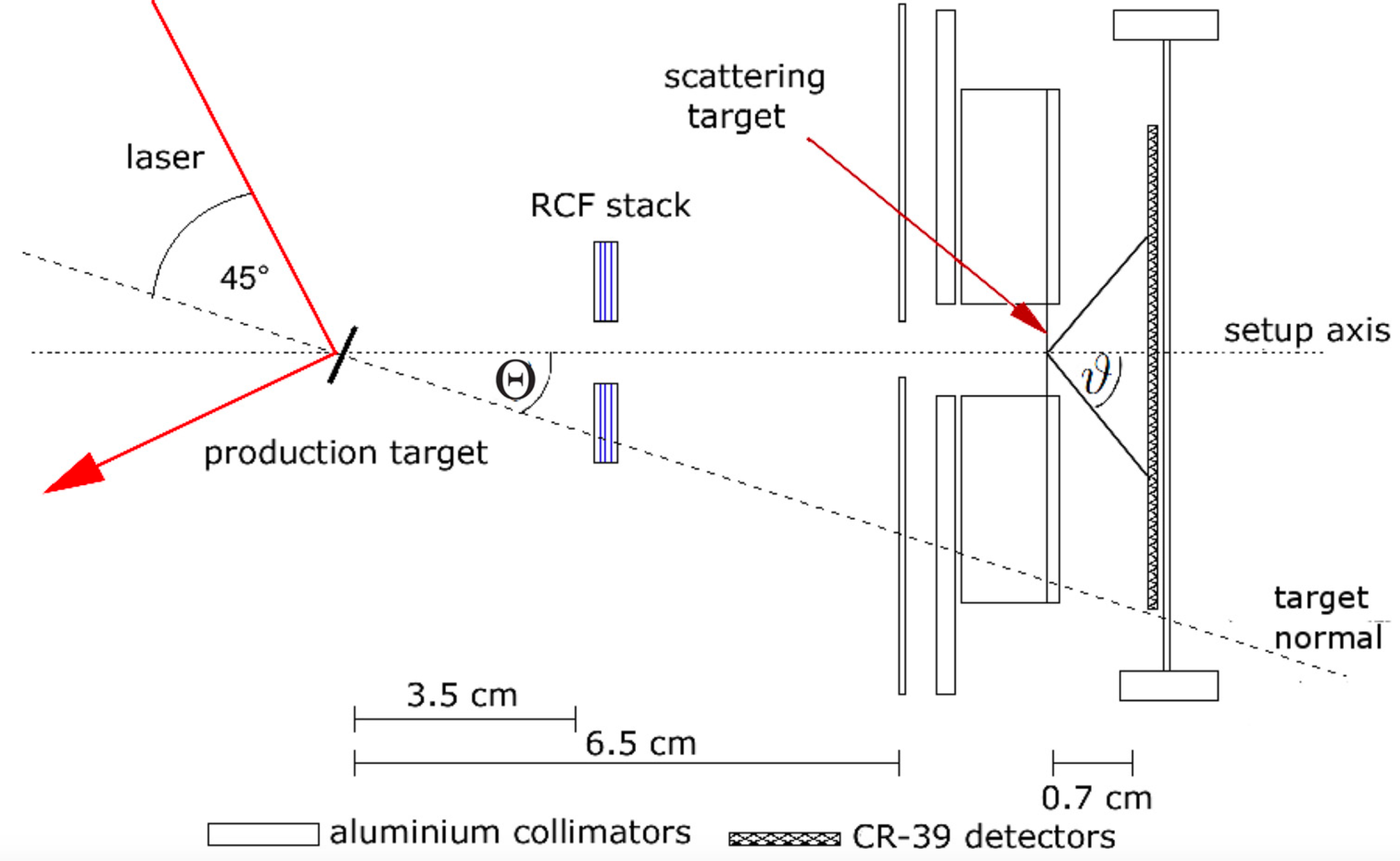}
\caption{Schematic setup for the first proton polarization measurement. \cite{bib2}}
\end{figure} 

\setlength{\parindent}{0 mm}  
\hspace*{5 mm}
To estimate the magnitude of possible polarizing magnetic fields in this case, Particle-In-Cell (PIC) simulations have been carried out with the fully relativistic 2D code EPOCH \cite{bib2, bib7}. A \textit{B}-field strength of $\sim$\,$10^{4}$\,T and gradients of $10^{10}$\,Tm$^{–1}$ are expected. Although these values are rather high, they are yet too small to align the proton spins and do not yield of measurable proton polarization. \\ \hspace*{5 mm}
One conclusion from this experiment is that for measuring a proton polarization \textit{P}\,$\neq$\,0 both, a stronger laser pulse with an intensity of about $10^{23}$\,{Wcm$^{-2}$ and an extended gas instead of a thin foil target are needed. Such a scenario has been theoretically considered in a paper by Shen \textit{et al.} \cite{bib8}. Due to a larger target size, the interaction time between the laser accelerated protons and the \textit{B}-field is increased. The typical time scale for spin motion is given by the Larmor frequency. For the numbers in Ref.\,[8] this is in the order of $0.1$\,ps, i.e., sufficiently short compared to the interaction time of approximately $3.3$\,ps of the accelerated protons with the magnetic field and, thus, a spin manipulation is possible. \\ \hspace*{5 mm}
With respect to gas targets it has been demonstrated that for nuclear and electron spin-polarized hydrogen at densities of at least $\sim$\,10$^{19}$\,cm$^{-3}$ the polarization lifetime is $\sim\,$10\,ns which is sufficiently long to generate polarized hydrogen atoms on the timescale of our experiment \cite{bib9}. This density is large enough for laser-driven ion acceleration of spin-polarized protons.

\section{PROTON SPIN DYNAMICS}
We have implemented particle-spin effects into the 3D PIC simulation code VLPL (Virtual Laser Plasma Lab) in order to make theoretical predictions about the degree of proton-spin polarization from a laser-driven plasma accelerator \cite{bib10, bib11}. These calculations consider all relevant effects that may lead to the polarization of proton beams \cite{bib12}.
\\ \hspace*{5 mm}
The Sokolov-Ternov effect is, for example, employed in classical accelerators to polarize the stored electron beams where the typical polarization build-up times are minutes or longer. This effect can, therefore, be neglected in the case of laser-induced acceleration. We refer to our forthcoming publication Ref.\,[12] for a more quantitative estimate. \\ \hspace*{5 mm}
Our assessment for the Stern-Gerlach force \cite{bib12} shows that non-relativistic proton beams with opposite spins are separated by not more than $\Delta_{p}$\,$\approx$ 9.3$\cdot$10$^{-7}\,\lambda_{\mathrm{L}}$ with the laser wavelength $\lambda_{\mathrm{L}}$. Moreover, the field strengths is of the order of $E$$\,\approx\,$\textit{B}$\,\approx$\,10$^{5}$\,T and the field gradients $\nabla|\textbf{B}|$\,$\approx$\,10$^{5}$\,\textit{T}/\textit{R} with the laser radius \textit{R}, typically $\lambda_{\mathrm{L}}$/\textit{R}\,=\,1/10 and a characteristic separation time would be \textit{t}\,=\,100\,$\omega_{\mathrm{L}}^{-1}$ where $\omega_{\mathrm{L}}$ is the laser frequency. Thus, the force on the given length scale is too weak and the Stern-Gerlach effect does not have to be taken into account for further simulation work on proton-spin tracking.

\hspace*{5 mm}
For charged particles the spin precession in arbitrary electric and magnetic fields is given by the T-BMT equation \cite{bib13} in CGS units

\begin{equation}
\centering
\begin{aligned}
\frac{d\textbf{s}}{dt}=-\frac{e}{m_{\mathrm{p}}c} \bigg\lbrack \bigg( a_{\mathrm{p}}+ \frac{1}{\gamma} \bigg) \textbf{B} - \frac{a_{\mathrm{p}}\gamma}{\gamma+1} \bigg( \frac{\textbf{v}}{c} \cdot \textbf{B} \bigg) \frac{\textbf{v}}{c} \\ - \bigg( a_{\mathrm{p}} + \frac {1}{1+\gamma} \bigg ) \frac{\textbf{v}}{c} \times \textbf{E} \bigg\rbrack \times \textbf{s}= - \vec{\Omega} \times \textbf{s} \, \, .
\end{aligned} 
\end{equation}

Here \textbf{s} is the proton spin in the rest frame of the proton, \textit{e} is the elementary charge, $m_{\mathrm{p}}$ the proton mass, \textit{c} the speed of light, the dimensionless anomalous magnetic moment of the proton $a_{\mathrm{p}}$\,=\,$\frac{g_{\mathrm{p}}-2}{2}$\,=\,1.8 with the \textit{g}-factor of the free proton $g_{\mathrm{p}}$, $\gamma$ the Lorentz factor, \textbf{v} the particle velocity, \textbf{B} the magnetic field, and \textbf{E} the electric field, both in the laboratory frame. Since $\vec{\Omega}$ always has a component perpendicular to \textbf{s}, the single spins in a polarized particle ensemble precess with the frequency $\omega_{\mathrm{s}}$\,=\,$|\vec{\Omega}|$. For protons with an energy in the range of a few GeV, $\gamma$\,$\approx$\,1 and 1\,$\gtrsim$\,$\textbf{v}/c$, so that
\begin{equation}
\centering
\begin{aligned}
\omega_{\mathrm{s}} < \frac{e}{m_{\mathrm{p}} c} \sqrt{(a_{\mathrm{p}}+1)^{2}\,\textbf{B}^{2}+\bigg(\frac{a_{\mathrm{p}}}{2}\bigg)^{2}\,\textbf{B}^{2}+\bigg(a_{\mathrm{p}}+\frac{1}{2}\bigg)^2\,\textbf{E}^{2}} \, \, .
\end{aligned} 
\end{equation} 
Under the assumption $|\textbf{B}|$\,$\approx$\,$|\textbf{E}|$\,$\approx F$ this simplifies to
\begin{equation}
\centering
\begin{aligned}
\omega_{\mathrm{s}} < \frac{e}{m_{\mathrm{p}} c} F \sqrt{\frac{9}{4}\, a_{\mathrm{p}}^{2}+3 \, a_{\mathrm{p}}+\frac{5}{4}} \, \, .
\end{aligned} 
\end{equation}
As a consequence, a conservation of the polarization of the system is expected for times
\begin{equation}
\centering
\begin{aligned}
t \ll \frac{2\pi}{\omega_{\mathrm{s}}}\approx \frac{2\pi}{3.7 \frac{e}{m_{\mathrm{p}} c} F}
\end{aligned} 
\end{equation}
for $a_{\mathrm{p}}$\,=\,1.8. For typical field strengths in our performed simulations (cf. Fig.\,3.1) of \textit{F}\,=\,5.11\,$\cdot10^{12}$\,V/m\,=\,17.0\,$\cdot10^{3}$\,T the preservation of the spin directions is estimated for times \textit{t}\,$<$\,1\,ps. This time is sufficiently long taking into account that the simulation time is $t_{\mathrm{sim}}$\,=\,0.13\,ps\,$\ll$\,1\,ps, so the polarization is maintained during the entire simulation according to the T-BMT equation.

\section{PARTICLE-IN-CELL SIMULATIONS}
In order to reproduce the results of the seminal experiment presented in Sec.\,1 \cite{bib2} and to verify the quantitative estimates of Ref.\,[2], 3D simulations with the above mentioned VLPL code including spin tracking have been carried out on the supercomputer JURECA \cite{bib14}. These were performed for a focused 3D laser pulse of Gaussian shape with wavelength $\lambda_{\mathrm{L}}$\,=\,800\,nm, a normalized laser amplitude \textit{a}$_{0}$\,=\,12 calculated for the ARCturus laser system, a duration of 25\,fs and a focal spot size of 5\,µm. \\ \hspace*{5 mm}
It is important to consider that to simulate the plasma behavior, a PIC code first defines a three-dimensional Cartesian grid which fills the simulated volume where the plasma evolves over the simulated time. Moreover, not each physical particle is treated individually but they are substituted by so-called PIC particles. This is why the continuous spin vector of a PIC particle represents the mean spin of all substituted particles. Thus, not the spin of each single particle is simulated but the polarization \textit{P} of every PIC particle. Therefore, the sum of spin vectors of different PIC particles within a certain volume (polarization cell) corresponds to the local polarization of the ensemble \cite{bib12, bib15}.
\hspace{-10cm}
\begin{figure*}
	\begin{subfigure}[c]{0.5\textwidth}
		\centering
		\includegraphics[width=0.97\textwidth]{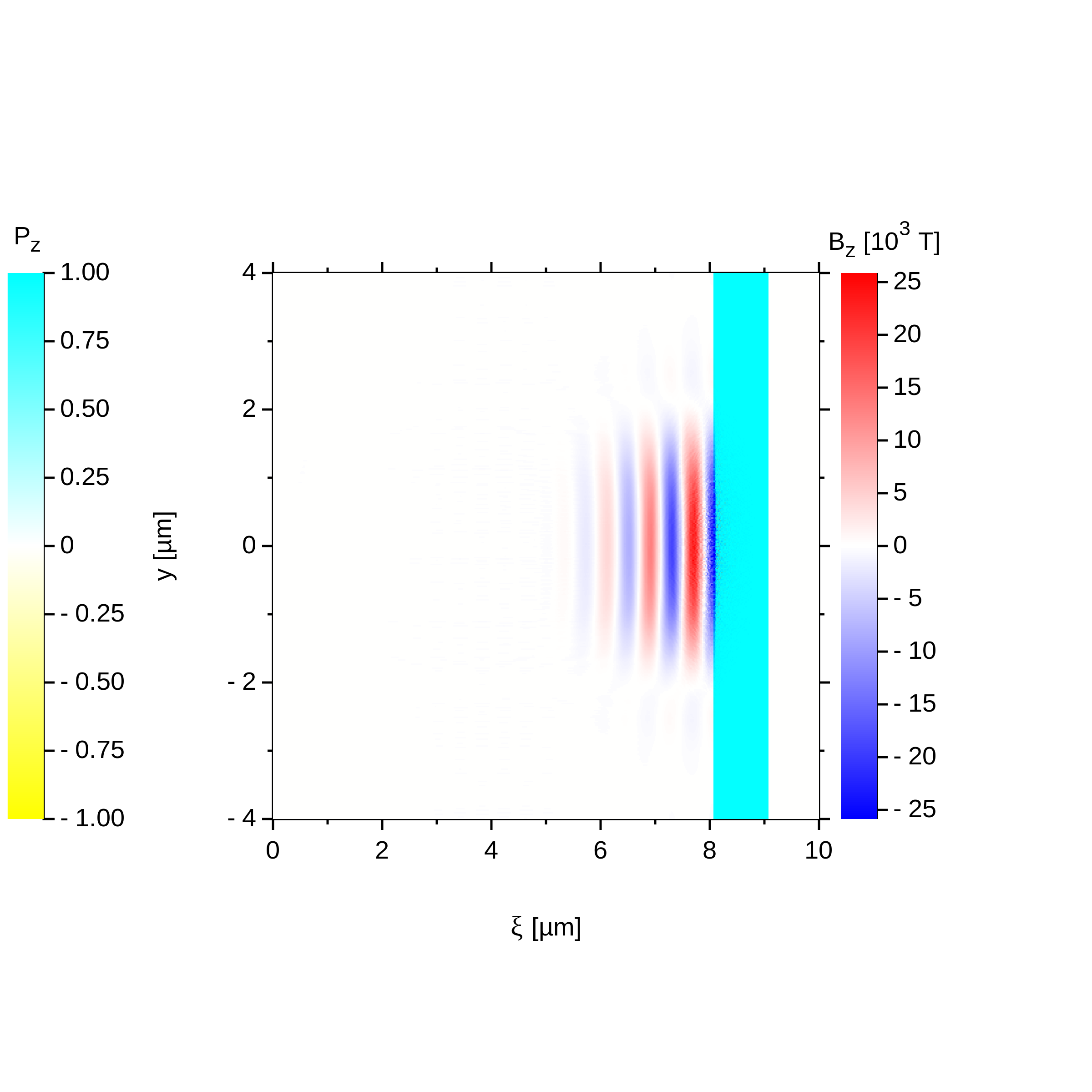}
		\subcaption[width=0.9\textwidth]{Hydrogen target (thickness 1\,µm, density 128\,$n_\mathrm{cr}$) at simulation time 13\,fs. $P_\mathrm{z}$ indicates the polarization and $B_\mathrm{z}$ the magnetic field.}
	\end{subfigure}
	\begin{subfigure}[c]{0.5\textwidth}
		\centering
		\includegraphics[width=0.97\textwidth]{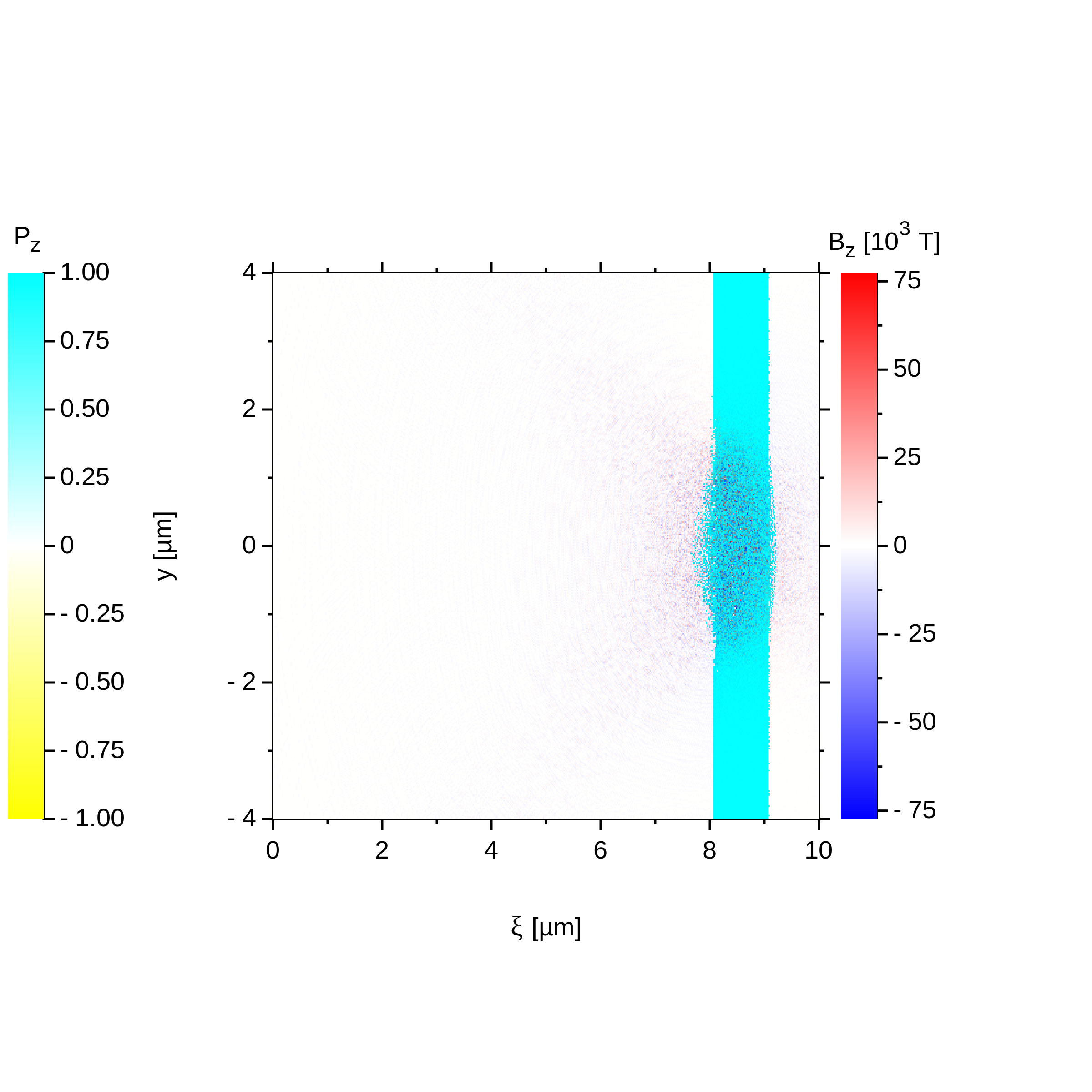}
		\subcaption[width=0.9\textwidth]{Same as a) at simulaton time 65\,fs. The proton polarization is preserved within the entire interaction process.}
	\end{subfigure}
\hspace{-10cm}
	\begin{subfigure}[c]{0.5\textwidth}
	\centering
	\includegraphics[width=0.97\textwidth]{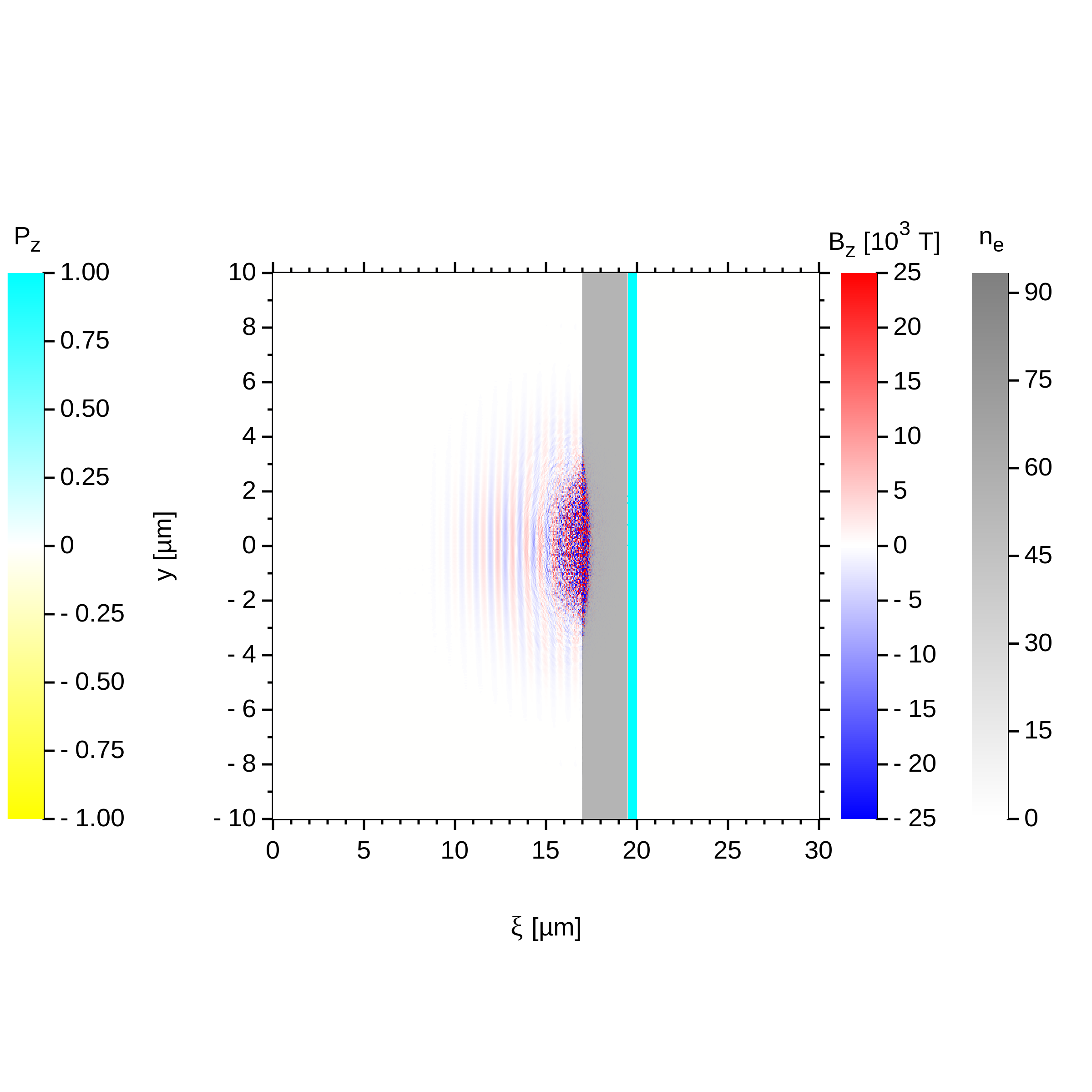}
	\subcaption{Aluminum foil target (2.5\,µm, 35\,$n_\mathrm{cr}$) covered with a fully polarized proton layer (0.5\,µm, 117\,$n_\mathrm{cr}$) at simulation time 32.5\,fs. $n_\mathrm{e}$ represents the electron density.} 
	\end{subfigure}
	\begin{subfigure}[c]{0.5\textwidth}
	\centering
	\includegraphics[width=0.97\textwidth]{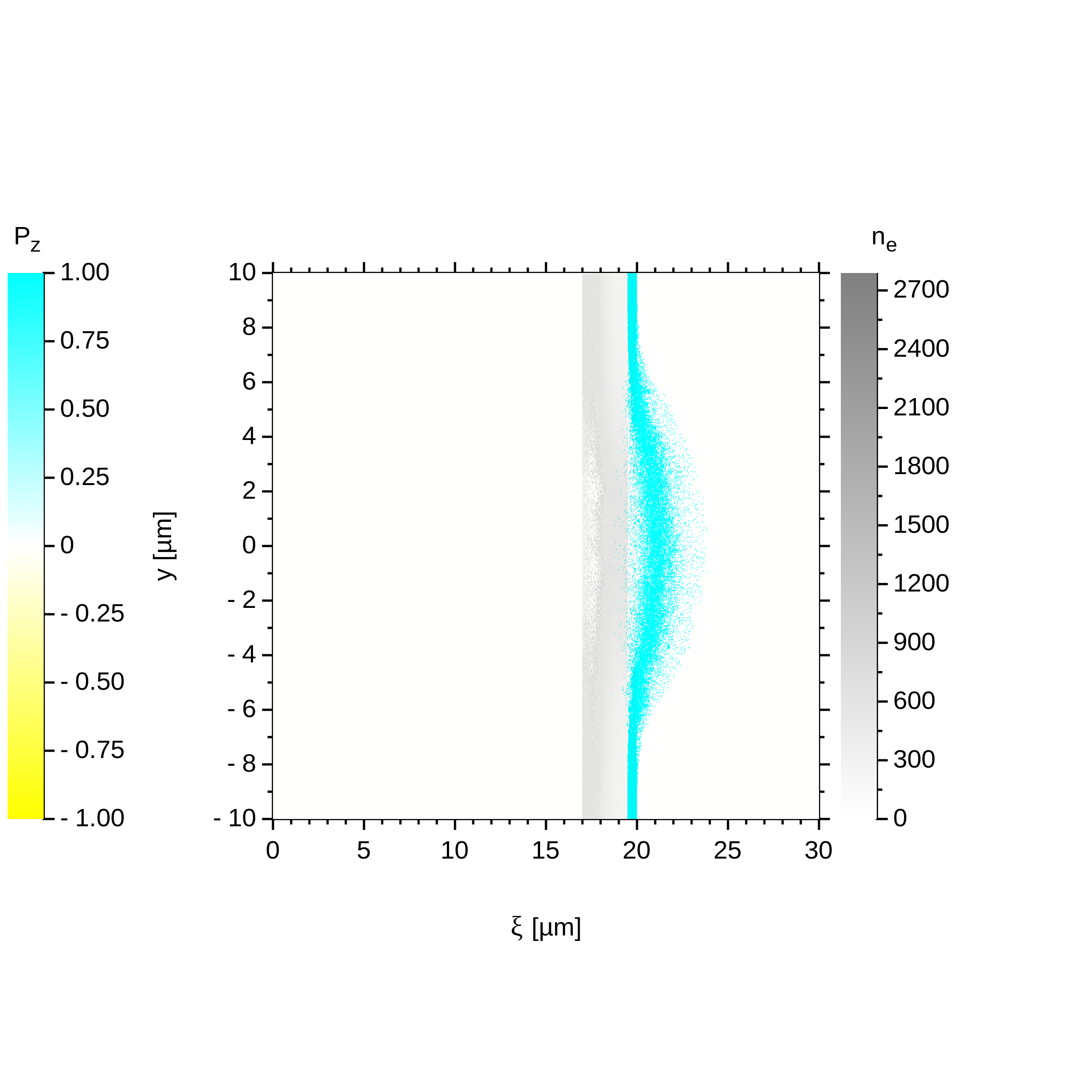}
	\subcaption{Same as c) at simulation time 130\,fs. The proton polarization is preserved both within the interaction process and within the acceleration process.} \hspace{3cm}
	\end{subfigure}
\caption{3D VLPL simulations showing the conservation of proton polarization in two polarized target geometries after interaction with a laser pulse ($\lambda_{\mathrm{L}}$\,=\,800\,nm, normalized laser amplitude \textit{a}$_{0}$\,=\,12, 25\,fs duration, 5\,µm focal spot size) impinging from the left side of the simulation box.}
\end{figure*}
\enlargethispage{\baselineskip}
\\ \hspace*{5 mm}
Figure 3.1 shows preliminary simulation results for proton-spin tracking with the PIC code VLPL. Two different simulation scenarios were investigated regarding the development of proton spins in the interaction with a laser pulse. For this purpose, the simulations were carried out with many particles per cell and a fully polarized hydrogen layer. \\ \hspace*{5 mm}
The upper two images depict the magnetic field $B_\mathrm{z}$ and the polarization $P_\mathrm{z}$ distribution for a pure hydrogen target (thickness 1\,µm, density 128\,$n_\mathrm{cr}$). For the simulation a grid cell size of $h_\mathrm{x}$\,=\,$h_\mathrm{y}$\,=\,$h_\mathrm{z}$\,=\,0.02\,µm was chosen. Within the target geometry the polarization is preserved after interaction with the laser pulse, impinging from the left side of the simulation box. The resulting field strengths are in the range of 7.5\,$\cdot$\,10$^{4}$\,T, so one can assume that the polarization is preserved for up to 0.24\,ps. \\ \hspace*{5 mm}
In the lower two pictures a more complicated scenario is chosen, which is very close to the setup described in Sec.\,1. The laser impinges on a an Aluminum foil target (2.5\,µm, 35\,$n_\mathrm{cr}$) covered with a fully polarized proton layer (0.5\,µm, 117\,$n_\mathrm{cr}$). A grid cell size $h_\mathrm{x}$\,= 0.025\,µm and $h_\mathrm{y}$\,=\,$h_\mathrm{z}$\,=\,0.05\,µm was used. An acceleration of the protons due to the TNSA mechanism is in evident. The fields that interact in the target here are more static and we estimate a proton polarization preservation for at least 0.18\,ps.\\ \hspace*{5 mm}
Thus, VLPL simulations on proton polarization demonstrate the conservation of polarization according to the T-BMT equation when accelerated by the TNSA mechanism \cite{bib13,bib15}. Our analysis of the spin-rotation angle in the simulations shows a precession of most PIC particles by less than 15 degrees which confirms the conservation of polarization. Considering that, a compact target is needed in which the nuclear spins are already aligned at the time of irradiation with the accelerating laser. For an in-depth analysis of particle acceleration with polarized targets, we refer to Ref.\,[12] which will be published shortly. However, solid foil targets suitable for laser acceleration with TNSA mechanism are not available so far and an experimental realization is extremely challenging. In previous experiments hydrogen nuclear polarization mostly results from a static polarization, e.g., in frozen spin targets \cite{bib16} or with polarized $^{3}$He gas \cite{bib17, bib18}. For the acceleration of protons until now only polarized atomic beam sources based on the Stern-Gerlach principle \cite{bib19} are available, which however have the disadvantage of a too small particle density. In order to provide a dynamically polarized hydrogen gas target for laser-plasma applications, a new approach is needed.

\section{EXPERIMENTAL REALIZATION}
For the experimental realization of our new concept for a dynamically polarized ion source, three components are required: a suitable laser system, a vacuum interaction chamber including a gas jet and a polarimeter. The schematic view of the setup is depicted in Fig.\,4.1.\\ \hspace*{5 mm}
As a component of the gas target, hydrogen halides are a viable option \cite{bib20,bib21}. A hydrogen chloride (HCl) target is preferred in this case due to the rather high polarizability and the easy availability. The HCl gas is injected into the interaction chamber by a standard gas nozzle with a high-speed short-pulse piezo valve that can be operated at 5\,bar inlet-gas pressure to produce a gas density in the range of $\sim$\,10$^{19}$\,cm$^{-3}$. Few millimeters below the nozzle, the interaction between gas and laser beams takes place. \\ \hspace*{5 mm}
\begin{figure}[H]
	\centering
	\includegraphics[width=85mm]{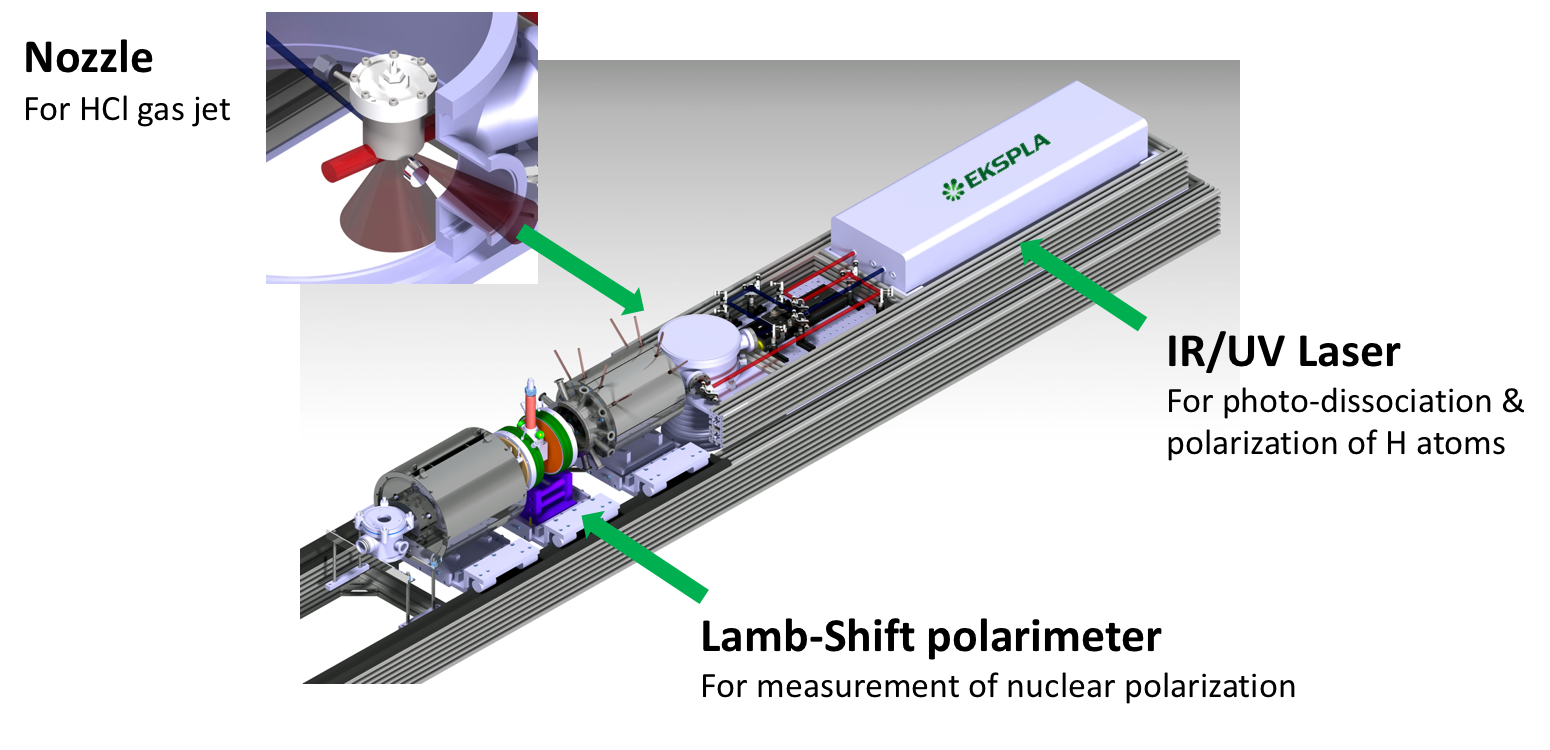}
	\caption{Schematic view of the setup for the proton polarization measurement using a polarized hydrogen gas target.}
\end{figure}
\hspace*{5 mm}
The polarizing laser system is a pulsed Ni:YAG laser from EKSPLA \cite{bib22}. Its peculiarity is the quasi-simultaneous output of the fundamental wavelength at 1064\,nm and the fifth (213\,nm) harmonic. The repetition rate of the laser system is 5\,Hz and the pulses are of 170\,ps duration which is sufficiently short with regard to the transfer time of the electron spin polarization to the nucleus due to hyperfine interaction ($\sim$\,1\,ns) \cite{bib20}. The linearly polarized 1064\,nm beam with a pulse energy of 100\,mJ is focused with an intensity of $\sim$\,10$^{11}$\,Wcm$^{-2}$ into the interaction chamber to align the HCl bonds (cf.\,Fig.\,4.2). By this, the signal intensity is increased and the amplification factor \textit{x} is calculated to be \textit{x}\,$\approx$\,2 assuming an interaction parameter of $\Delta\omega$\,=\,49 and, thus, $\ll $cos$^{2} \theta$$\gg$\,=\,6/7 since the polarizability interaction is governed by a cos$^{2} \theta$ potential with the angle $\theta$ between the molecular axis and the electric field distribution \cite{bib23}. \\ \hspace*{5 mm} 
At the same time but under a 90$^{\circ}$ angle, the circularly polarized fifth harmonic with an energy of 20\,mJ is also focused at an intensity of $\sim$\,10$^{12}$\,Wcm$^{-2}$ into the vacuum chamber to interact with the HCl gas. The aligned HCl molecules are photo-dissociated by UV excitation via the $\textit{A}^{1}\mathrm{\Pi}_{1}$ state, which has a total electronic angular-momentum projection of $\Omega$\,=\,+1 along the bond axis. Hence, the resulting H and Cl($^{2}$P$_{3/2}$) photofragments conserve this +1 projection of the laser photons, producing H and Cl($^{2}$P$_{3/2}$) atoms each with the projections of approximately $m_{\mathrm{s}}$\,=\,+1/2 (so that they sum to +1), and thus the H-atom electron spin is approximately $m_{\mathrm{s}}$\,=\,+1/2 \cite{bib24}. In a weak magnetic field (Zeeman region), all H atoms are in a coherent superposition of the total angular momentum states $|F,m_{\mathrm{F}} \rangle$ with the coupling $\textbf{F}$\,=\,$\textbf{S}$\,+\,$\textbf{I}$ of the electron spin \textbf{S} and the nuclear spin \textbf{I}. When the electron spin is fixed due to the polarization of the incident laser beam, e.g., $m_{\mathrm{s}}$\,=\,+1/2, then only the spin combinations  $|m_{\mathrm{s}}$\,=\,+1/2,  $m_{\mathrm{I}}$\,=\,+1/2$>$ and $|$+1/2,\,-1/2$>$ can be found in the free hydrogen atoms. The hyperfine state $|$+1/2,\,+1/2$>$\,=\,$|F$\,=\,1, $m_\mathrm{F}$\,=\,+1$>$ is an eigenstate and will stay unchanged in time. Since the states $|$-1/2,\,+1/2$>$ and $|$+1/2,\,-1/2$>$ are not eigenstates, they will be expressed as linear combinations of the eigenstates $|F$\,=\,1, $m_\mathrm{F}$\,=\,0$>$ and $|F$\,=\,0,\,$m_\mathrm{F}$\,=\,0$>$, which have different energies. Therefore, atoms produced in the $|$+1/2,\,-1/2$>$ state will oscillate to the $|$-1/2,\,+1/2$>$ state and back. If now the electron-polarized hydrogen atoms are produced during a very short time \textit{t}\,$<$\,1\,ns, they will oscillate in phase. Therefore, after 0.35\,ns only the spin combinations $|$+1/2,\,+1/2$>$ and $|$-1/2,\,+1/2$>$ are found. This means that the electron polarization of the hydrogen atoms, produced by the laser beam, is transferred into a nuclear polarization. If now the hydrogen atoms are ionized and accelerated, the out-coming protons will remain polarized, even if they undergo spin precessing according to the T-BMT equation \cite{bib20}.

\begin{figure}[H]
	\centering
	\includegraphics[width=85mm]{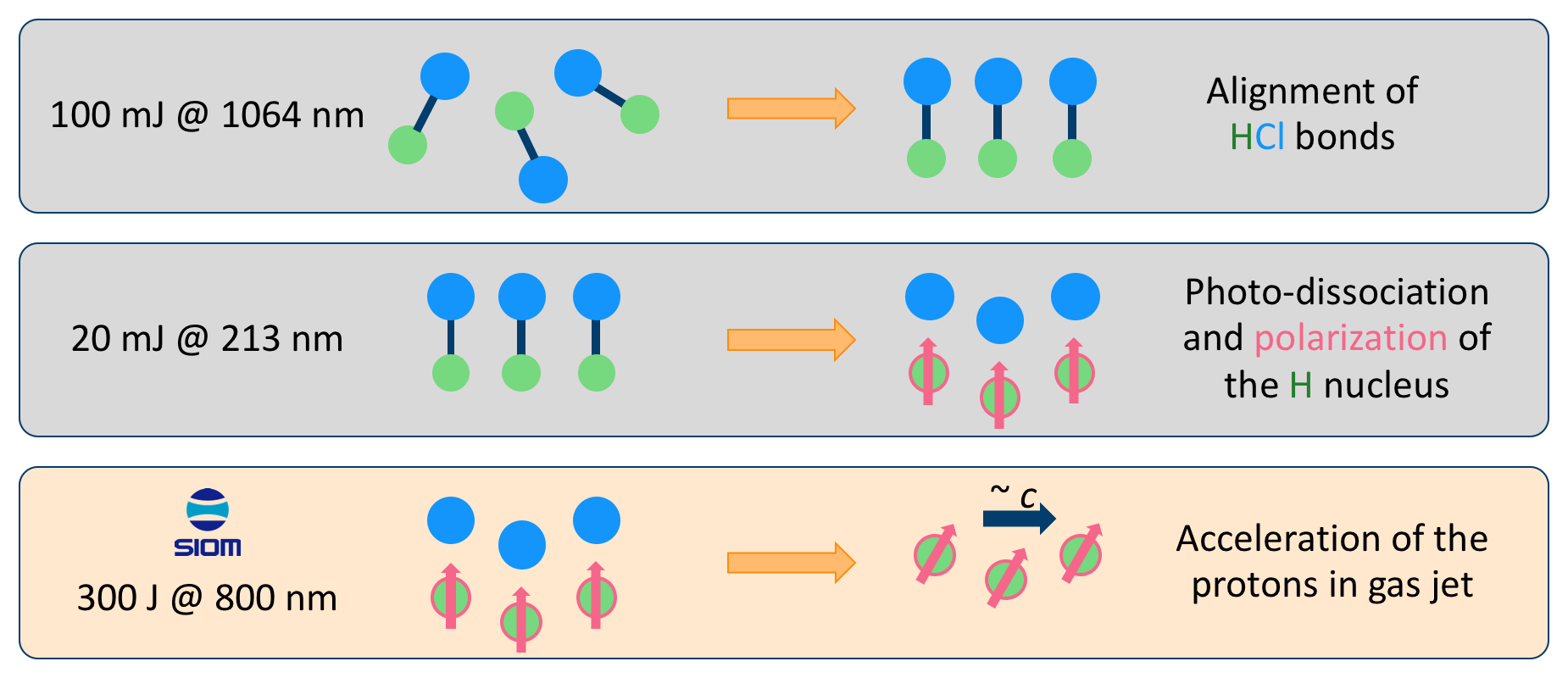}
	\caption{Schematic overview of the production of polarized proton beams.}
\end{figure}
\setlength{\parindent}{0 mm}  
\hspace*{5 mm}
Using a Lamb-Shift polarimeter the polarization of an atomic hydrogen ensemble can be measured in a multi-step process \cite{bib25,bib26}. One important condition is that the atomic beam can be efficiently converted into metastable atoms in the 2S$_{1/2}$ state by ionization with an electron-impact ionizer and a charge reversal in cesium vapor. With a spin filter, individual hyperfine sub-states are selected by applying a static magnetic field, an electric quench field and a high-frequency transition. By varying the resonance condition when changing the magnetic field, single hyperfine components can be detected. Finally, the transition into the ground state within the quenching process is verified by Lyman-$\alpha$ radiation emitted at 121.5\,nm. The intensity of the individual hyperfine components allows to measure their occupation number and, therefore, calculate the polarization of incoming protons and in combination with an ionizer even for hydrogen atoms. The entire setup, including laser system, interaction chamber and Lamb-Shift polarimeter, is realized over a length of less than 5\,m as a table-top experiment. \\ \hspace*{5 mm}
To summarize, our novel gas target will offer nuclear polarized hydrogen atoms at a density of 10$^{19}$\,cm$^{-3}$ or above with a one-to-one mixture of (unpolarized) chlorine atoms. The suitability of such type of target, i.e., containing hydrogen and an admixture of heavier nuclei, for proton acceleration has already been demonstrated with the help of PIC simulations (although without considering spin effects) in Ref.\,[27]. It was found that laser intensities of $>$\,$10^{22}$\,Wcm$^{-2}$ promise to reach proton energies above 1 GeV. Such a laser system will be available in the near future at the Shanghai Institute of Optics and Fine Mechanics (SIOM). The Shanghai Superintense-Ultrafast Lasers Facility (SULF) will offer pulse energies of 300\,J at 30\,fs pulse duration and a repetition rate of 1\,shot/min.  Another important conclusion from Ref.\,[27] is that the heavy ions are not accelerated from the gas target, however, they are vital to provide the proton acceleration in a so-called electron bubble-channel structure. In this acceleration scheme protons, which are trapped in the bubble region of the wake field, can be efficiently accelerated in the front of the bubble, while electrons are mostly accelerated at its rear. After the acceleration process the proton polarization will be determined by a detector similar to that one described in Sec.\,1.

\section{DISCUSSION AND CONCLUSION}
In conclusion, the T-BMT equation, describing the spin precession in electromagnetic fields, has been implemented into the VLPL PIC code to simulate the spin behavior during laser-plasma interactions. One crucial result of our simulations is that a target containing polarized hydrogen nuclei is needed for producing polarized relativistic proton beams. A corresponding gas-jet target, based on dynamic polarization of HCl molecules, is now being built at Forschungszentrum Jülich. By interacting the fundamental wavelength of a Nd:YAG laser and its fifth harmonic with HCl gas, nuclear polarized H atoms are created. Their nuclear polarization will be measured and tuned with a Lamb-shift polarimeter. First measurements, aiming at the demonstration of the feasibility of the target concept, are scheduled for fall 2018. The ultimate experiment will take place at the 10\,PW SULF facility to observe a up to GeV polarized proton beam from laser-generated plasma for the first time.

\section{ACKNOWLEDGMENT}
We thank our colleagues B.\,F. Shen, L.\,Ji, J.\,Xu and L.\,Zhang from Shanghai Institute of Optics and Fine Mechanics for the various fruitful discussions and their expertise that greatly assisted our research. This work has been carried out in the framework of the \textit{Ju}SPARC (Jülich Short-Pulse Particle and Radiation Center) project and has been supported by the ATHENA (Accelerator Technology HElmholtz iNfrAstructure) consortium. We further acknowledge the computing resources on grant VSR-JPGI61 on the supercomputer JURECA.

\appendix{}

\end{document}